\documentclass[sigconf, 10pt]{acmart}
\pdfoutput=1
\renewcommand\footnotetextcopyrightpermission[1]{} 
\def\BibTeX{{\rm B\kern-.05em{\sc i\kern-.025em b}\kern-.08emT\kern-.1667em\lower.7ex\hbox{E}\kern-.125emX}}

\setcopyright{none}

\usepackage{graphicx}
\usepackage{multirow}
\usepackage{subcaption}
\usepackage{stfloats}
\usepackage{algorithm}
\usepackage{algpseudocode}
\begin{document}

\settopmatter{printacmref=false}

%
\title{Graph Sampling with Distributed In-Memory Dataflow Systems}

%
\author{Kevin Gomez}
\email{gomez@informatik.uni-leipzig.de}
\affiliation{%
  \institution{University of Leipzig}
}

\author{Matthias T{\"a}schner}
\email{taeschner@informatik.uni-leipzig.de}
\affiliation{%
  \institution{University of Leipzig}
}

\author{M. Ali Rostami}
\email{rostami@informatik.uni-leipzig.de}
\affiliation{%
  \institution{University of Leipzig}
}

\author{Christopher Rost}
\email{rost@informatik.uni-leipzig.de}
\affiliation{%
  \institution{University of Leipzig}
}

\author{Erhard Rahm}
\email{rahm@informatik.uni-leipzig.de}
\affiliation{%
  \institution{University of Leipzig}
}

%
\renewcommand{\shortauthors}{Gomez, et al.}

%
\begin{abstract}
Given a large graph, a graph sample determines a subgraph with similar characteristics for certain metrics of the original graph. The samples are much smaller 
thereby accelerating and simplifying the analysis and visualization of large graphs. 
We focus on the implementation of distributed graph sampling for Big Data frameworks and in-memory dataflow systems such as Apache Spark or Apache Flink. We evaluate the scalability of the new implementations and analyze to what degree the sampling approaches preserve certain graph metrics compared to the original graph. The latter analysis also uses comparative graph visualizations. 
The presented methods will be open source and be integrated into  \textsc{Gradoop}, a system for distributed graph analytics.
\end{abstract}

%
%


%

%
\maketitle

\section{Introduction}
\label{sec.intro}
Sampling  is used to determine  a subset of a given dataset that retains certain properties but allows more efficient data analysis. For graph sampling it is necessary to retain not only general characteristics of the original data but also the structural information. 
 Graph sampling is especially important for the efficient processing and analysis of large graphs such as social networks \cite{Leskovec:2006:SLG:1150402.1150479,5961350}. Furthermore, sampling is often needed to allow the effective visualization of large graphs. 
 
 Our contribution in this paper is to outline the distributed implementation of known graph sampling algorithms for improved scalability to large graphs as well as their evaluation. The sampling approaches are added as operators to the open-source distributed graph analysis platform \textsc{Gradoop}\footnote{http://www.gradoop.com} \cite{junghanns2016epgm,Junghanns2018DeclarativeAD} and used for interactive graph visualization \cite{rostami_2019}.
Our distributed sampling algorithms are, like \textsc{Gradoop},  based on the  dataflow execution framework Apache Flink but the implementation would be similar for Apache Spark. The evaluation  for different graphs considers the runtime scalability as well as the quality of sampling regarding retained graph properties and the similarity of graph visualizations.   




This paper is structured as follows: We briefly discuss  related work in Section~\ref{sec.related} and provide background information on graph sampling  in Section~\ref{sec.background}. In  Section~\ref{sec.impl}, we explain the distributed implementation of four sampling algorithms with Apache Flink. Sections~\ref{sec.eval} describes the evaluation results before we conclude in Section ~\ref{sec.conc}.

\section{Related Work}
\label{sec.related}
Several previous publications address graph sampling algorithms but mostly without considering their distributed implementation. 
Hu et al.~\cite{Hu2013ASA} survey different graph sampling algorithms and their evaluations. However, many of these algorithms cannot be applied to large graphs due to their complexity. Leskovec et al.~\cite{Leskovec:2006:SLG:1150402.1150479}  analyze sampling algorithms for large graphs but there is no discussion of distributed or parallel approaches. Wang et al.~\cite{5961350} focuses on  sampling algorithms for social networks but again without considering distributed approaches. 

The only work about distributed graph sampling we are aware of is a recent paper by Zhang et al.~\cite{Zhang:2018:2470-1173:379} for implementations based on  Apache Spark. In contrast to our work, they do not evaluate the speedup behavior for different  cluster sizes and  the scalability to different data volumes.  Our study also includes a distributed implementation and  evaluation of random walk sampling.

\section{Background}
\label{sec.background}

We first introduce some basic definition of a graph sample and a graph sample algorithm. Afterwards, we specify some basic sampling algorithms and outline important graph metrics for both, visual and metrical comparison in the evaluation chapter.

\subsection{Graph Sampling} 

A directed graph $G=(V,E)$ can be used to express the interactions of users of a social network. The user  can be denoted as a vertex $v \in V$ and a relationship between two user $v$ and $u$ can be denoted as a directed edge $e=(v,u) \in E$.

Since popular social networks such as Facebook and Twitter contains billions of users and trillions of relationships, the resulting graph is too big for both, visualization and analytical tasks. A common approach to reduce the size of the graph is to use graph sampling to scale down the information contained in the original graph.

\textbf{Definition 1} (\textsc{Graph Sample}) A graph $S = (V_S, E_S)$ is a sampled graph (or graph sample) of graph $G=(V,E)$ iff the following three constraints are met: $V_S \subseteq V$, $E_S \subseteq E$ and $E_S \subseteq \{(u,v)|u \in V_S, v\in V_S\}$.\footnote{In the existing publications, there are different approaches toward the vertices with zero-degrees in the sampled graph. Within this work we choose the approach to remove all zero-degree vertices from the sampled graph.}

\textbf{Definition 2} (\textsc{Graph Sample Algorithm}) A graph sample algorithm is a function from a graph set $\mathcal{G}$ to a set of sampled graphs $\mathcal{S}$, as $f:\mathcal{G}\to \mathcal{S}$ in which the set of vertices $V$ and edges $E$ will be reduced until a given threshold $s \in [0,1]$ is reached. $s$ is called \textit{sample size} and defines the ratio of vertices (or edges) the graph sample contains compared to the original graph.

A graph sample is considered to be fruitful if it can represent many properties of the original graph. For example, if we want to find dense communities, like a group of friends in a social network, the sampled graph is only worthwhile if it preserves these relations as much as possible. We evaluate this concept by comparing some predefined graph properties on both original and sampled graphs. 

\subsection{Basic Graph Sampling Algorithms}
Many graph sampling algorithms have already been investigated  but we will limit ourselves to four basic approaches in this paper: 
\textit{random vertex sampling}, \textit{random edge sampling}, \textit{neighborhood sampling}, and \textit{random walk sampling}. 

Random vertex sampling is the most straightforward sampling approach that uniformly samples the graph by selecting a subset of vertices and their corresponding edges based on the selected sample size $s$. For the distributed implementation in a shared-nothing approach, the information of the whole graph is not always available in every node. Therefore, we consider an estimation by selecting the vertices using $s$ as a probability. This approach is also applied on the edges in the random edge sampling.


The idea of the random neighborhood sampling is to improve topological locality over the simple random vertex approach. Therefore, when a vertex is chosen to be in the resulting sampled graph, all neighbors are also added to the sampled graph. Optionally, only incoming or outgoing edges can be taken into account to select the neighbors of a vertex. 

For the random walk sampling, one or more vertices are randomly selected as start vertices. For each start vertex, we follow a randomly selected outgoing edge to its neighbor. If a vertex has no outgoing edges or if all edges were followed already, we jump to any other randomly chosen vertex in the graph and continue the walk there. To avoid keeping stuck in dense areas of the graph we added a probability to jump to another random vertex instead of following an outgoing edge. This process continues until a desired number of vertices have been visited and the sample size $s$ has been met. All visited vertices and all edges whose source and target vertex was visited will be part of the graph sample result.

\subsection{Important Graph Metrics}
As we mentioned, we present an evaluation on the visual and metrical comparison of the original graph and the sampled one. Following is a set of graph metrics which will be evaluated in Section \ref{sec.eval}.
\begin{itemize}
    \item Cardinality of vertex and edge set, denoted as $|V|$ and $|E|$.
    \item Graph density $\mathcal{D}$: The ratio of all actually existing edges to all possible edges in the graph, defined by: $$\mathcal{D}=\frac{|E|}{|V|(|V|-1)}$$
    \item Number of triangles $\mathcal{T}$: The number of subsets of vertices in the graph with three elements which are fully connected (triangle or closed triple).  
    \item Global clustering coefficient $\mathcal{C_G}$: The ratio of the number of triangles $\mathcal{T}$ to the number of all triples $|Triple|$ in the graph (see~\cite{BOCCALETTI2006175}), defined by: $$\mathcal{C_G}=\frac{3\cdot\mathcal{T}}{|Triple|}$$  
    \item Average local clustering coefficient $\mathcal{C_L}$: The local clustering coefficient of a vertex is the ratio of the edges that actually connect its neighbors to the maximum possible number of edges between those neighbors. With a value between $0$ and $1$, it describes how close the vertex and its neighborhood are to a clique (see~\cite{watts1998smallworldnetworks}). We compute the average local clustering coefficient for all vertices in the graph.
    \item Number of weakly connected components $|WCC|$: A maximal connected subgraph of a graph in which each two vertices can be reached through a path is called a weakly connected component. The number of such components in a graph is the target parameter.
    \item The average, minimum and maximum vertex degree in the graph, denoted as $d_{avg}$, $d_{min}$, and $d_{max}$.
\end{itemize}

\section{Implementation}
\label{sec.impl}

The goal of the distributed  implementation of graph sampling are to achieve fast execution and good scalability for large graphs with up to billions of vertices and edges. We therefore want to utilize the parallel processing capabilities of shared-nothing clusters and, specifically, distributed dataflow systems such as Apache Spark~\cite{zaharia2012resilient} and Apache Flink~\cite{carbone2015apache}. In contrast to the older MapReduce approach, these frameworks offer a wider range of transformations and keep data in main memory between the execution of operations. Our  implementations are based on Apache Flink but can be easily transferred to Apache Spark. We first give a brief introduction to the programming concepts of the distributed dataflow model. We then outline the implementation of our sampling operators.

\subsection{Distributed Dataflow Model}

The processing of data that exceeds the computing power or storage of a single computer can be handled through the use of distributed dataflow systems. Therein the data is processed simultaneously on shared-nothing commodity cluster nodes. Although details vary for different frameworks, they are designed to implement parallel data-centric workflows, with datasets and primitive transformations as two fundamental programming abstractions. A \textit{dataset} represents a typed collection partitioned over a cluster. A \textit{transformation} is a deterministic operator that transforms the elements of one or two datasets into a new dataset. A typical distributed program consists of chained transformations that form a dataflow. A scheduler breaks each dataflow job into a directed acyclic execution graph, where the nodes are working threads and edges are input and output dependencies between them. Each thread can be executed concurrently on an associated dataset partition in the cluster without sharing memory. 

Transformations can be distinguished into \textit{unary} and \textit{binary} operators, depending on the number of input datasets. Table~\ref{tab:transformations} shows some common transformations from both types which are relevant for this work.
The \textit{filter} transformation evaluates a user-defined predicate function to each element of the input dataset. If the function evaluates to \texttt{true}, the element is part of the output. Another simple transformation is \textit{map}. It applies a user-defined map function to each element of the input dataset which returns exactly one element to guarantee a one-to-one relation to the output dataset. A transformation processing a group instead of a single element as input is \textit{reduce} where the input, as well as output, are key-value pairs. All elements inside a group share the same key. The transformation applies a user-defined function to each group of elements and aggregates them into a single output pair. A common binary transformation is \textit{join}. It creates pairs of elements from two input datasets which have equal values on defined keys. A user-defined join function is applied for each pair that produces exactly one output element.

\begin{table}[]
    \centering
    \begin{tabular}{llll}
    \toprule
       \textbf{Transf.}  & \textbf{Type} & \textbf{Signature} & \textbf{Constraints} \\ \midrule
       Filter   & unary     & $I,O\subseteq A$              & $O\subseteq I$ \\
       Map      & unary     & $I\subseteq A, O\subseteq B$  & $|I|=|O|$ \\
       Reduce   & unary     & $I,O\subseteq A\times B$      & $|I|\geq |O|\wedge |O|\leq |A|$\\
       Join     & binary    & $O\subseteq I_1\Join I_2$     & $I_1\subseteq A, I_2\subseteq B$ \\
      \bottomrule
    \end{tabular}
    (I/O : input/output datasets, A/B : domains)
    \caption{Selected transformations}
    \label{tab:transformations}
\end{table}

\subsection{Sampling Operators}

The operators for  graph sampling  compute a subgraph by either randomly selecting a subset of vertices or a subset of edges. In addition, neighborhood information or graph traversal can be used. The computation uses a series of transformations on the input graph. Latter is stored in two datasets, one for vertices and one for edges. For each sampling operator a filter is applied to the output graph's vertex dataset to remove all zero-degree vertices following the definition of a graph sample in Section \ref{sec.background}.

\subsubsection{\underline{R}andom \underline{V}ertex (\textbf{RV}) and \underline{R}andom \underline{E}dge (\textbf{RE}) Sampling}

The input for the RV and RE operator is a input graph and a sample size $s$. For RV, a filter operator is applied to the vertex dataset of the input graph. A vertex will be kept if a generated random value $r \in [0,1]$ is lower or equal to $s$. An edge will be kept if its source and target vertices occur in the dataset of the remaining vertices.

\begin{figure*}
\centering
\begin{minipage}{\textwidth}
	\begin{minipage}{0.5\textwidth}
    	\centering    	
    	\includegraphics[width=\linewidth]{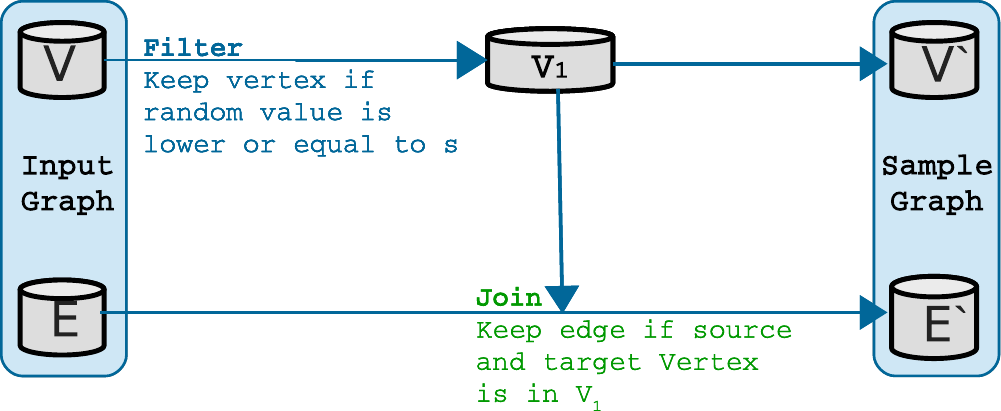}
    	\caption{Dataflow RV Operator.}
    	\label{fig:vertex-sampling-dataflow}
  	\end{minipage}	
	\begin{minipage}{0.5\textwidth}
    	\centering
    	\includegraphics[width=\linewidth]{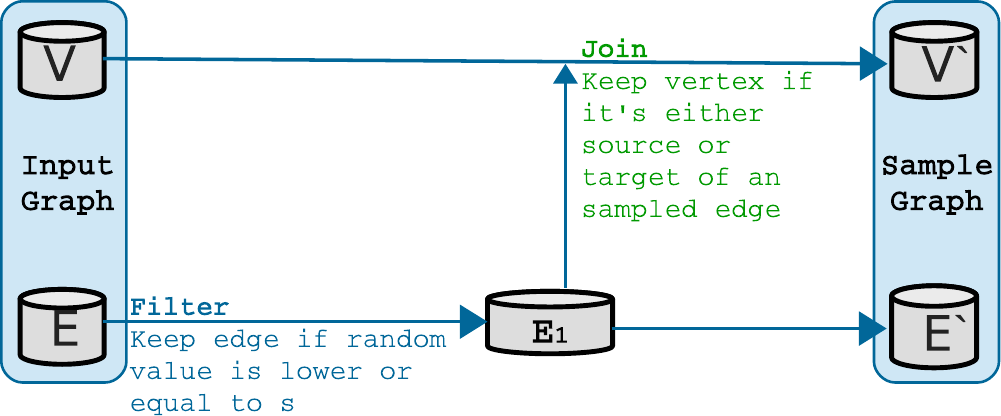}
    	\caption{Dataflow RE Operator.}
    	\label{fig:edge-sampling-dataflow}
  	\end{minipage}
\end{minipage}
\end{figure*}

RE works the other way around, as a filter transformation is applied to the edge dataset of the input graph. An edge will be kept, again if the generated random value $r \in [0,1]$ is lower or equal to $s$. A vertex will be kept if it's either the source or the target vertex of a remaining edge. The dataflow of the RV and RE operator can be seen in Figure \ref{fig:vertex-sampling-dataflow} and \ref{fig:edge-sampling-dataflow}. 


\subsubsection{\underline{R}andom \underline{V}ertex \underline{N}eighborhood (\textbf{RVN}) Sampling} 

This approach is similar to the RV operator but also adds the direct neighbors of a vertex to the sample. The selection of the neighbors can be restricted according to the direction of the connecting edge (incoming, outgoing or both). In the implementation, randomly selected vertices of the input vertex dataset are  marked as sampled with a boolean flag. As for RV, we select a vertex by setting the flag to true, if a generated random value $r \in [0,1]$ is lower or equal than the given sample size $s$ or set it to false otherwise. In a second step, the marked vertices are joined with the input edge dataset, transforming each edge into a tuple containing the edge itself and the boolean flags for its source and target vertex. A filter operator is applied to the edge tuples, retaining only those edges, whose source or target vertices where sampled and matching the given neighborhood relation. This relation will be either a neighbor on an incoming edge of a sampled vertex, a neighbor on an outgoing edge, or both. The dataflow from an input logical graph to an sampled graph is illustrated in Figure \ref{fig:rvn-dataflow}.

\subsubsection{\underline{R}andom \underline{W}alk (\textbf{RW}) Sampling}

\begin{figure*}
\centering
\begin{minipage}{\textwidth}
	\begin{minipage}{0.5\textwidth}
    	\centering    	
    	\includegraphics[width=\linewidth]{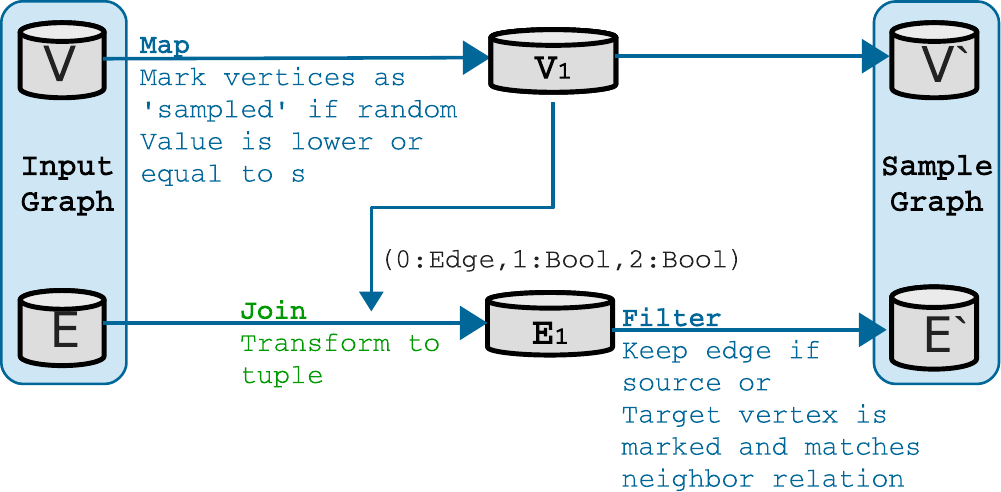}
    	\caption{Dataflow RVN Operator.}
    	\label{fig:rvn-dataflow}
  	\end{minipage}	
	\begin{minipage}{0.5\textwidth}
    	\centering
    	\includegraphics[width=\linewidth]{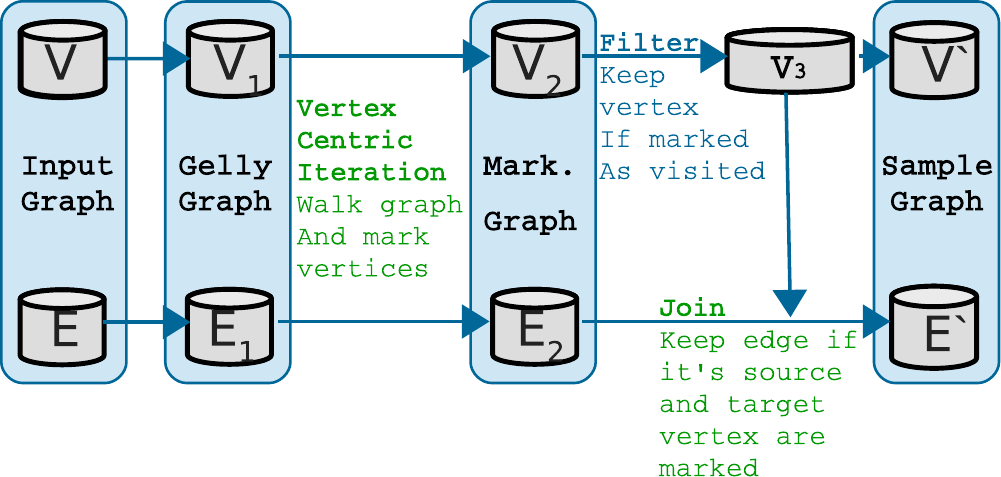}
    	\caption{Dataflow RW Operator.}
    	\label{fig:rw-dataflow}
  	\end{minipage}
\end{minipage}
\end{figure*}

This approach uses a random walk algorithm to walk over vertices and edges of the input graph. Each visited vertex and edges connecting those vertices will then be returned as the sampled graph. Figure \ref{fig:rw-dataflow} shows the dataflow of an input graph to a sampled graph of this operator. At the beginning we transform the input graph to a specific Gelly format. We are using Gelly, the Google Pregel~\cite{Malewicz:2010:PSL:1807167.1807184} implementation of Apache Flink, to implement a random walk algorithm.  

Pregel utilizes the bulk-synchronous-parallel~\cite{Valiant:1990:BMP:79173.79181} paradigm to create the vertex-centric-programming model. An iteration in a vertex-centric program is called \textit{superstep}. During a superstep each vertex of the graph can compute a new state in a \textit{compute} function. In a \textit{message} function each vertex is able to prepare messages for other vertices. At the end of each superstep each worker of the cluster can exchange the prepared massages during a synchronization barrier. In our operator we consider a message from one vertex to one of its neighbors a 'walk'. A message to any other vertex is considered as 'jump'.





At the beginning of the random walk algorithm a single start vertex is randomly selected and marked as visited. The marked vertex will be referred to as \textit{walker}. In the first superstep the walker either randomly picks one of its outgoing and not yet visited edges, walks to this neighbor and marks the edge as traversed. Or, with the probability of $j \in [0,1]$ or if there aren't any outgoing edges left, jumps to any other randomly selected vertex in the graph. Either the neighbor or the randomly selected vertex will become the new walker and the computation starts again. For a multi walk, more than one start vertex can be selected, which allows us to execute multiple walks in parallel.

For each completed superstep the already visited vertices are counted. If this number exceeds the desired number of sampled vertices, the iteration is terminated and the algorithm converges. Having the desired number of vertices marked as visited, the graph is transformed back and a filter operator is applied to its vertex dataset. A vertex will be kept if it is marked as visited. An edge will be kept if its source and target vertices occur in the dataset of the remaining vertices.

\section{Evaluation}
\label{sec.eval}

One key feature of distributed shared-nothing systems is their ability to respond to growing data sizes or problem complexity by adding additional machines. Therefore, we evaluate the scalability of our implementations  with respect to increasing data volume and computing resources in the first part of this section. The second part will contain a more visual comparison of our sampling algorithms. We will show, that our implementation computes expressive, structure-preserving graph samples based on the graph properties introduced in Section~\ref{sec.background}.

\textbf{Setup.} The evaluations were executed on a shared-nothing cluster with 16 workers connected via 1 GBit Ethernet. Each worker consists of an Intel Xeon E5-2430 6 x 2.5 Ghz CPU, 48 GB RAM, two 4 TB SATA disks and runs openSUSE 13.2. We use Hadoop 2.6.0 and Flink 1.7.0. We run Flink with 6 threads and 40 GB memory per worker.

We use two types of datasets for our evaluation: synthetic graphs to measure scalability of the algorithms and real-world graphs to metrically and visually compare the sampled and original graph.

To evaluate the scalability of our implementations we use the LDBC-SNB data set generator~\cite{erling2015ldbc}. It creates heterogeneous social network graphs with a fixed schema. The synthetic graphs mimic structural characteristics of real-world graphs, e.g., node degree distribution based on power-laws and skewed property value distributions. Table~\ref{tab:ldbc} shows the three datasets used throughout the benchmark. In addition to the scaling factor (SF) used, the cardinality of vertex and edge sets as well as the dataset size on hard disk are specified. Each is stored in the Hadoop distributed file system (HDFS). The execution times mentioned later include loading the graph from HDFS, computing the graph sample and writing the sampled graph back to HDFS. We run three executions per setup and report the average runtimes.

In addition, we use three real-world graphs from the SNAP Datasets~\cite{leskovec2014snapsets}, \textit{ego-Facebook}, \textit{ca-AstroPh} and \textit{web-Google}, to evaluate the impact of a sampling algorithm on the graph metrics and thus on the graphs structure. 

\subsection{Scalability}

In many real-world use cases data analysts are limited in graph size for visual or analytical tasks. Therefore, we run each sampling algorithm with the intention to create a sampled graph with round about 100k vertices. The used sample size $s$ for each graph is contained in Table~\ref{tab:ldbc}. For the RW operator 3000 walker and a jump probability $j = 0.1$ where used.

\begin{figure*}
\begin{minipage}{\textwidth}
	\begin{minipage}{0.33\textwidth}
    	\centering    	
    	\includegraphics[width=\linewidth,keepaspectratio]{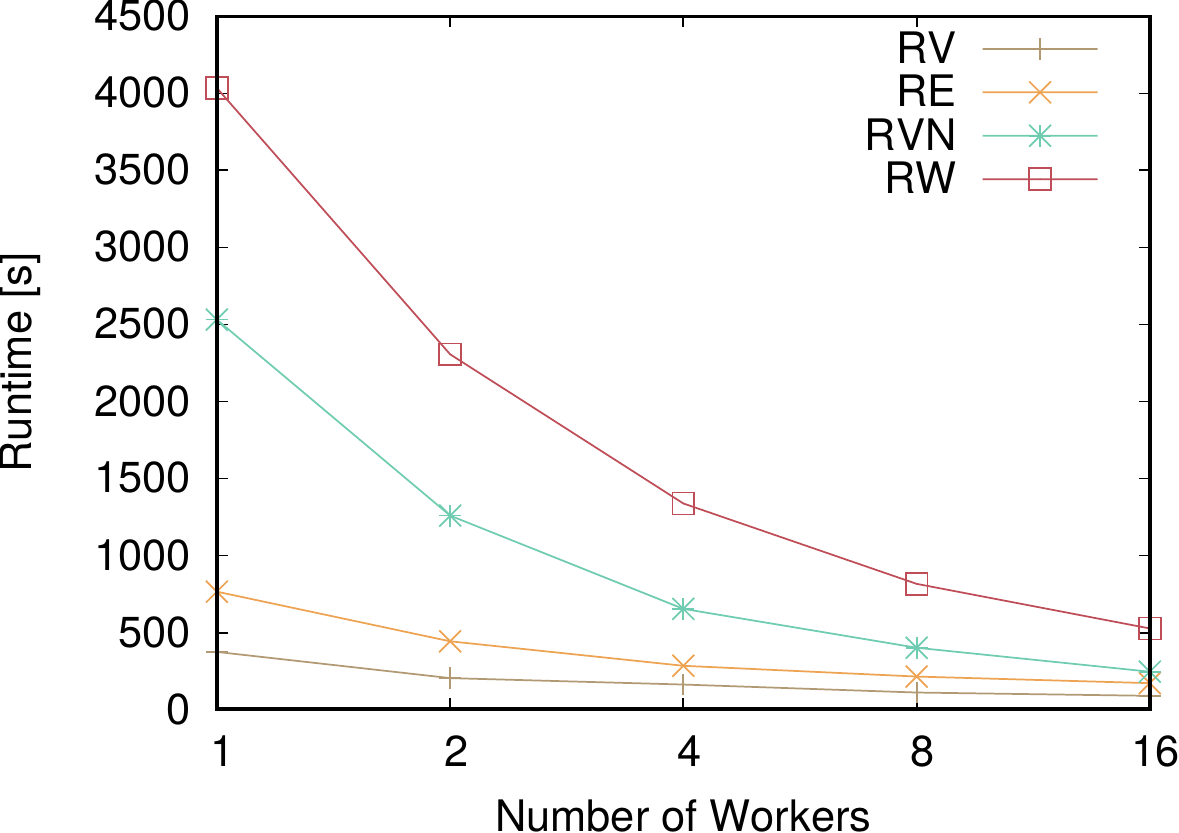}
    	\vspace{-8mm}
    	\caption{Increase worker count.}
    	\label{fig:eval:workercount}
  	\end{minipage}	
	\begin{minipage}{0.33\textwidth}
    	\centering
    	\includegraphics[width=\linewidth,keepaspectratio]{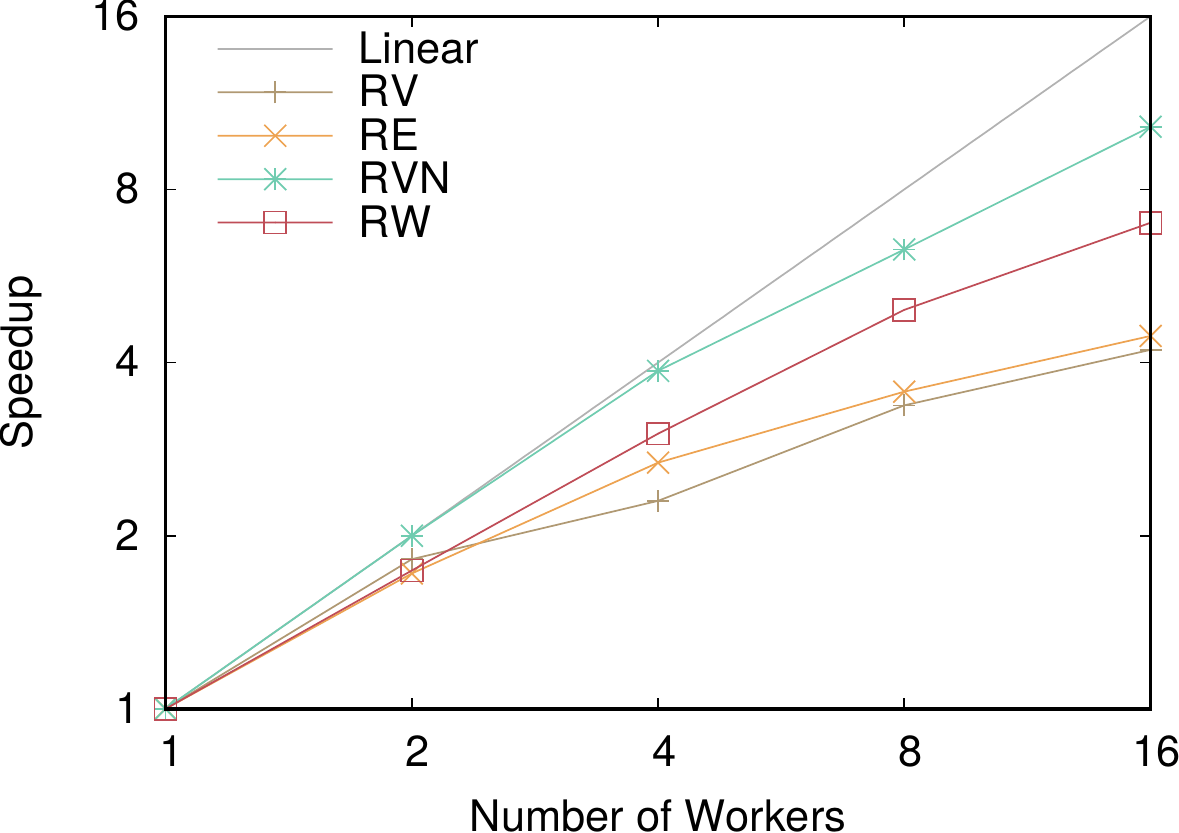}
    	\vspace{-8mm}
    	\caption{Speedup over workers.}
    	\label{fig:eval:speedup}
  	\end{minipage}
  	  	\begin{minipage}{0.33\textwidth}
    	\centering
    	\includegraphics[width=\linewidth,keepaspectratio]{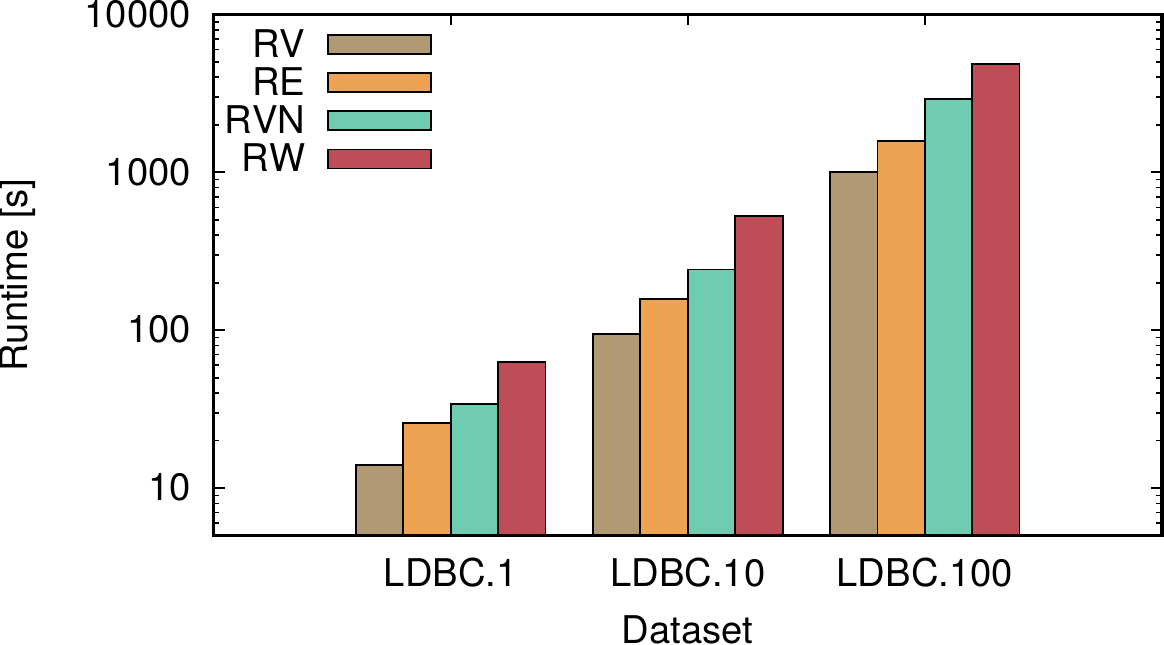}
    	\vspace{.5mm}
    	\caption{Increase data volume.}
    	\label{fig:eval:datasize}
  	\end{minipage}
\end{minipage}
\end{figure*}

We first evaluate the absolute runtime and relative speedup of our implementations.  Figure~\ref{fig:eval:workercount} shows the runtimes of the four algorithms for up to 16 workers using the \texttt{LDBC.10} dataset; Figure~\ref{fig:eval:speedup} shows the corresponding speedup values. While all algorithms benefit from more resources, \textit{RVN} and \textit{RW} gain the most. For \textit{RVN}, the runtime is reduced from 42 minutes on a single worker to 4 minutes on 16 workers (speedup 10.5). For \textit{RW}, a speedup of 7.45 is reached (reduction from 67 to 9 minutes). The simpler algorithms \textit{RV} and \textit{RE} are already executed fast on a single machine for \texttt{LDBC.10}. Hence, their potential for improvement is limited explaining the lower speedup values. 

\begin{table}[t]
    \centering
    \scriptsize
    \begin{tabular}{lrrrl}
        \toprule
        \multicolumn{1}{c}{\textbf{SF}}        & 
        \multicolumn{1}{c}{$|V|$}     & 
        \multicolumn{1}{c}{$|E|$}     & 
        \multicolumn{1}{c}{\textbf{Disk usage}} &
        \multicolumn{1}{c}{$s$}
        \tabularnewline
        \midrule
        1   & 3.3 M	 & 17.9 M  & 2.8 GB & 0.03
        \tabularnewline 
        \midrule
        10  & 30,4 M  & 180.4 M  & 23.9 GB & 0.003
        \tabularnewline 
        \midrule
        100 & 282.6 M  & 1.77 B   & 236.0 GB & 0.0003 
        \tabularnewline 
        \bottomrule
    \end{tabular}
    \caption{LDBC social network datasets}
    \label{tab:ldbc}
\end{table}



We also evaluate scalability with increasing data volume and a fixed number of workers (16 worker). The results in Figure~\ref{fig:eval:datasize} show that the runtimes of each algorithm increases almost linearly with growing data volume. For example, the execution of the \textit{RVN} algorithm required about 34 seconds on \texttt{LDBC.1} and 2907 seconds on \texttt{LDBC.100}. 

\subsection{Metric-Based and Visual Comparison}

An ideal sampling algorithm reduces the number of vertices and edges evenly by a desired amount. At the same time, the structural properties, as described by the calculated metrics, should be preserved in the best possible way. For example, a community of the graph can be thinned out, while the remaining vertices should stay equally connected to each other and thus hardly change their value for the local clustering coefficient.

In order to evaluate a sampling algorithm's impact on the graph metrics and thus on the graph structure, the metrics of an original and the sampled graph are compared. As mentioned before, we use three real-world graphs from the SNAP Datasets~\cite{leskovec2014snapsets}. Each sampling algorithm is applied three times to these graphs using a given sample size $s$ to reduce the number of vertices or edges by about 60\%. Due to the included neighborhood of each selected vertex, \textit{RVN} requires a much lower value for $s$ than the other sampling algorithms. For \textit{RW}, the number of walkers is scaled according to the number of vertices, starting with 5 walkers for the ego-Facebook graph, 20 walkers for the ca-AstroPh graph, and 1000 walkers for the web-Google graph. The jump probability $j = 0.1$ remained fixed throughout the experiment. We computed the proposed metrics for the original graphs and each resulting sample graph and added the average results to Table~\ref{tab:eval:metrics}. 

Since easier visualization is a main use case for graph sampling, we visually compare the original and the sampled graph structures for the ego-Facebook graph. Figure~\ref{fig:vis-eval:facebook-orig} shows the original graph with a force-directed layout~\cite{hu2005efficient}. The vertex size represents the degree, i.e. bigger vertices imply a higher degree. The vertex color stands for its local clustering coefficient, where a darker color represents a higher value. Figures~\ref{fig:vis-eval:facebook-rv}~to~\ref{fig:vis-eval:facebook-rw} show the sampled graphs for the different sampling algorithms. The positions of the vertices remain persistent compared to the original graph for all sampled graphs.

\textit{RV} manages to predictably reduce the number of vertices as well as the edges. According to Table~\ref{tab:eval:metrics}, the number of triangles in the graph has been reduced dramatically, the value of the global clustering coefficient is almost completely preserved. Depending on the original structure, \textit{RV} decomposes a graph into many new weakly connected components. As seen in Figure \ref{fig:vis-eval:facebook-rv}, \textit{RV} visibly thins out the graph but also destroys many of the existing communities and removes inter-community edges as well. \textit{RE} decreases the number of edges by the desired amount while hardly reducing the number of vertices. The value for the local clustering coefficient is reduced by a similar amount as the number of edges. All other structural properties of the original graph are unpredictably changed. The visualization in Figure \ref{fig:vis-eval:facebook-re} shows that most of the vertices are kept. The deleted edges reduce the connectivity within the communities and thus the local clustering coefficient of many vertices. \textit{RVN} reduces the number of vertices as desired and keeps about 5\% to 15\% of the edges.  Figure~\ref{fig:vis-eval:facebook-rvn} shows the well preserved neighborhood of sampled vertices, but the samples are lacking at edges connecting the individual communities. \textit{RW} reduces the number of vertices as expected. Figure~\ref{fig:vis-eval:facebook-rw} shows, that edges within the individual communities and edges connecting those communities tend to be preserved.

\section{Conclusion}
\label{sec.conc}

We outlined distributed implementations for four graph sampling approaches using Apache Flink. Our first experimental results are promising as they showed good speedup for using multiple workers and near-perfect scalability for increasing dataset sizes. The metric-based and visual comparisons with the original graphs confirmed that the implementations provide the expected, useful results thereby enabling the analyst and \textsc{Gradoop} user to select the most suitable sampling method. For example, both random vertex and random edge sampling are useful for obtaining an overview of the graph. Random vertex neighborhood (RVN) sampling is useful to analyze neighborhood relationships while random walk sampling is beneficial to study 
 the inter- and intra-connectivity of communities. In our ongoing work we provide distributed implementations for further  sampling algorithms such as Frontier Sampling and Forest Fire Sampling.

\section{Acknowledgments}

This work is partially funded by S{\"a}chsische Aufbau Bank (SAB) and
the European Regional Development (EFRE) under grant No. 100302179.

%
\bibliographystyle{ACM-Reference-Format}
\bibliography{main-bib}

\clearpage
\appendix
\section*{Appendix}

{

\begin{figure*}[bp]
\centering
	\begin{subfigure}{1.0\textwidth}
	    \centering
	    \includegraphics[width=0.35\linewidth]{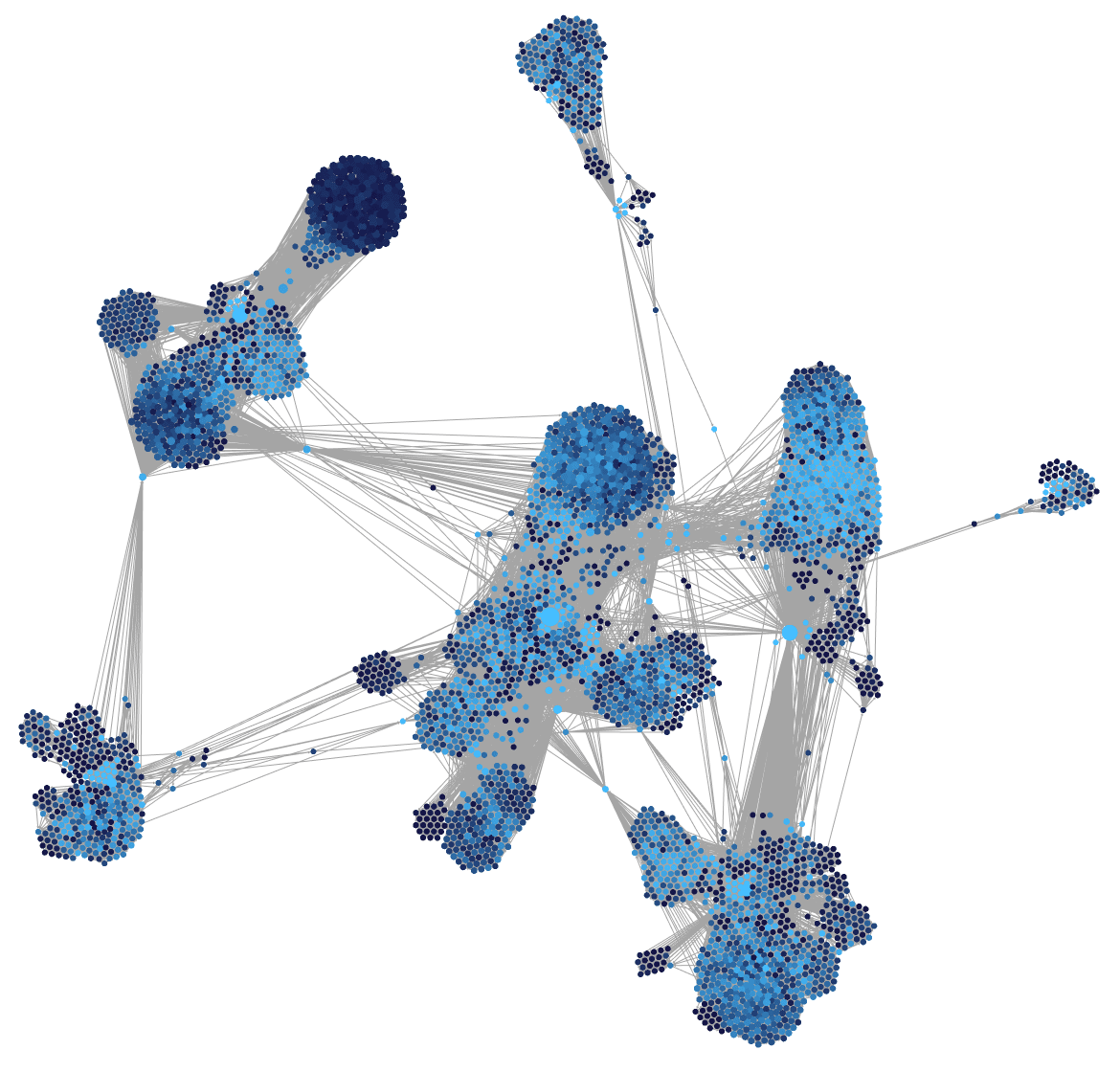}
        \caption{Original graph}
        \label{fig:vis-eval:facebook-orig}
	\end{subfigure}
    \begin{minipage}{0.8\textwidth}
        \centering
        \begin{subfigure}{0.45\linewidth}
            \centering
        	\includegraphics[width=\linewidth,keepaspectratio]{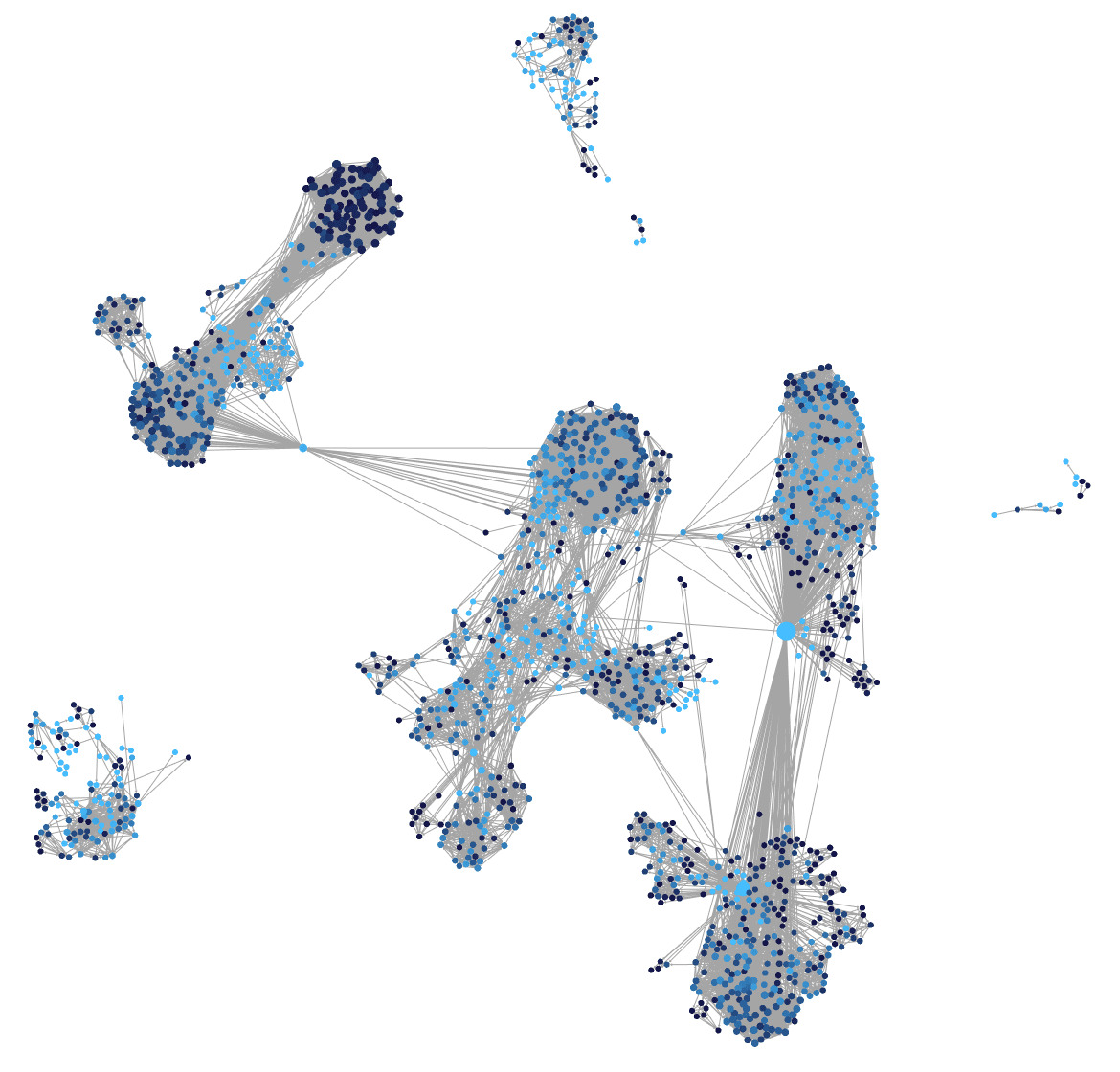}
        	\caption{\textit{RV}, $s = 0.4$}
        	\label{fig:vis-eval:facebook-rv}
        \end{subfigure}
        \hfill
        \begin{subfigure}{0.45\linewidth}
            \centering
        	\includegraphics[width=\linewidth,keepaspectratio]{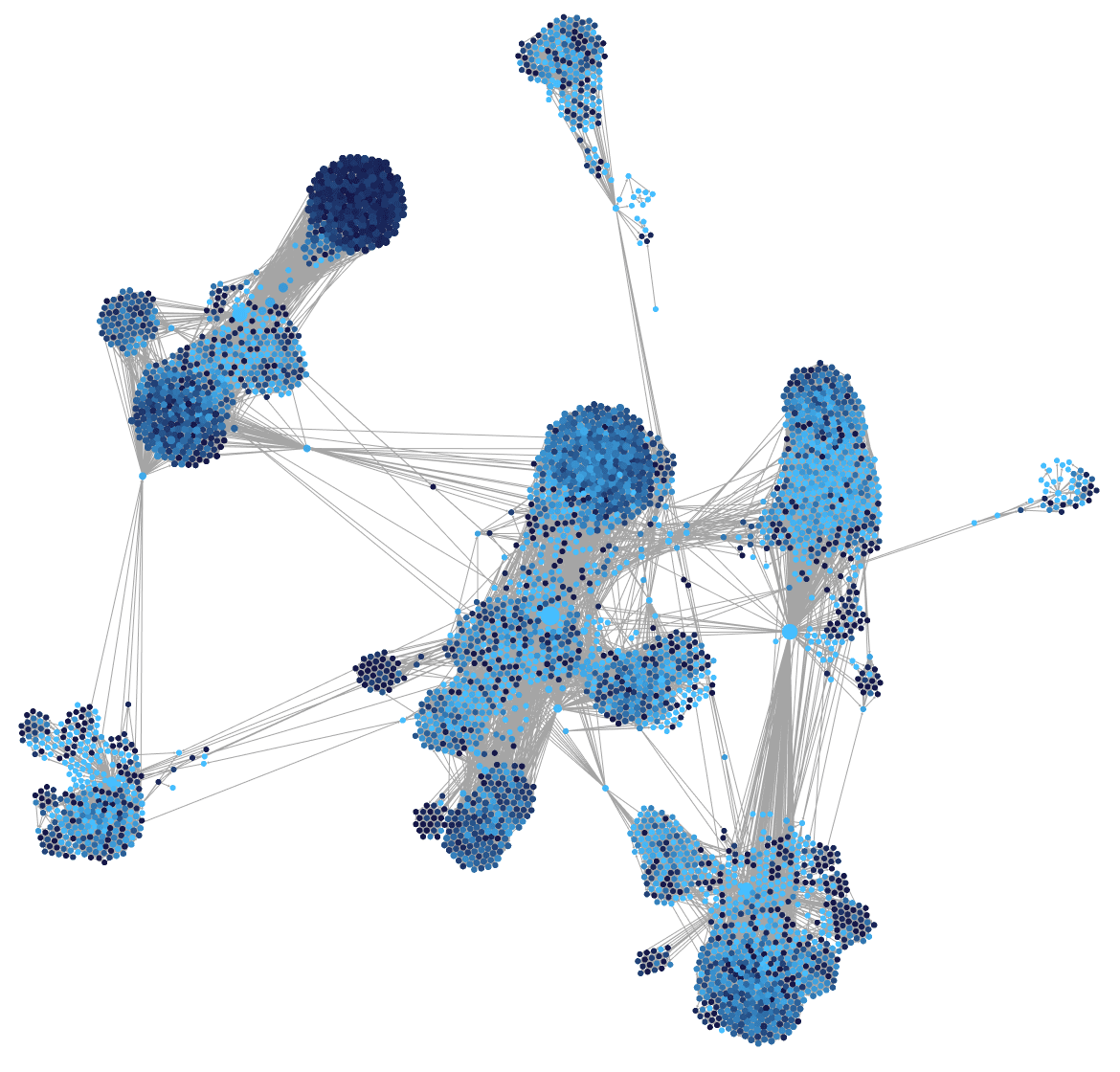}
        	\caption{\textit{RE}, $s = 0.4$}
        	\label{fig:vis-eval:facebook-re}
        \end{subfigure}
    \end{minipage}
	\begin{minipage}{0.8\textwidth}
    	\begin{subfigure}{0.45\linewidth}
    	    \centering
        	\includegraphics[width=\linewidth,keepaspectratio]{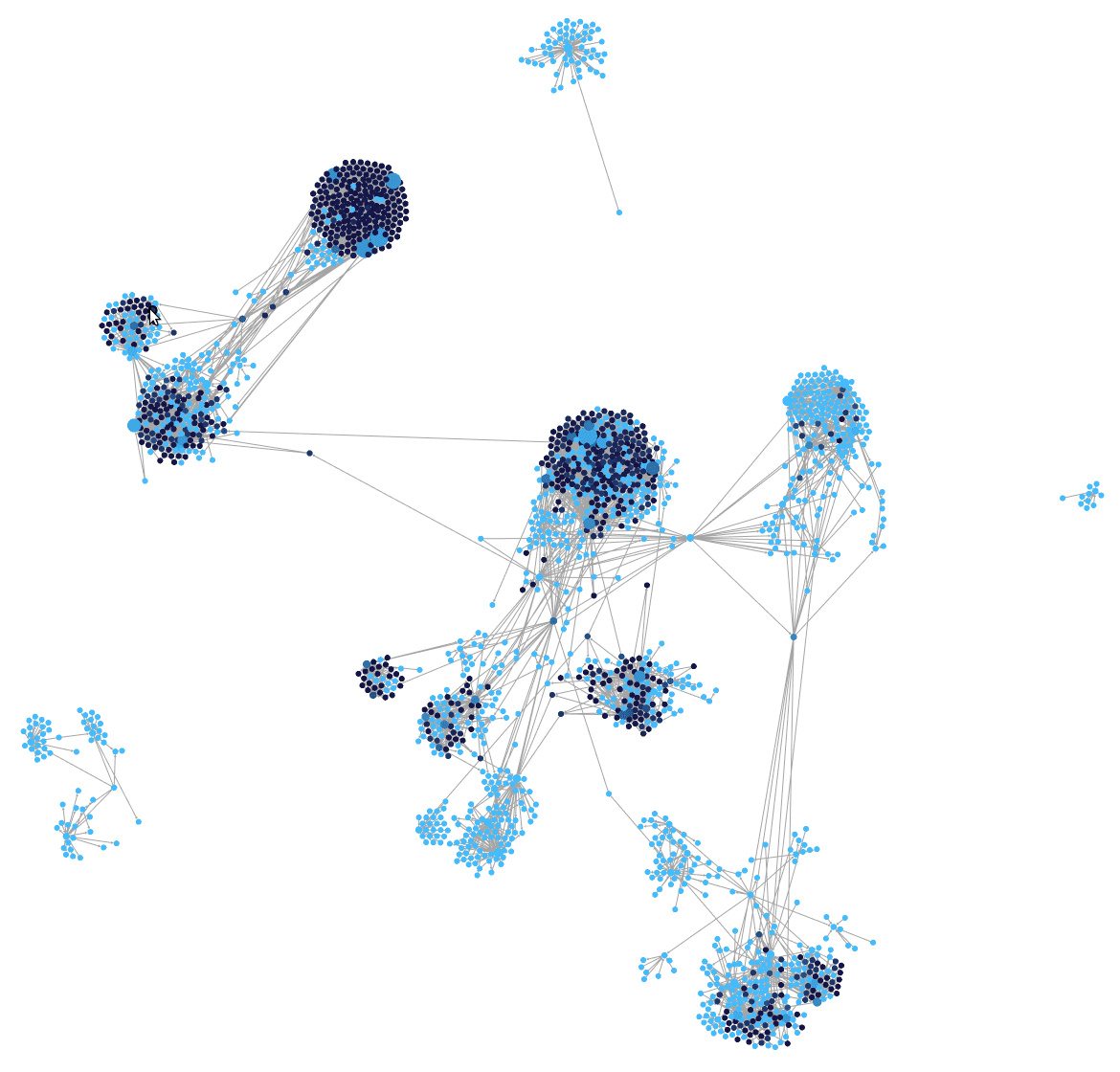}
        	\caption{\textit{RVN}, $s = 0.025$}
        	\label{fig:vis-eval:facebook-rvn}
    	\end{subfigure}
    	\hfill
    	\begin{subfigure}{0.45\linewidth}
    	    \centering
        	\includegraphics[width=\linewidth,keepaspectratio]{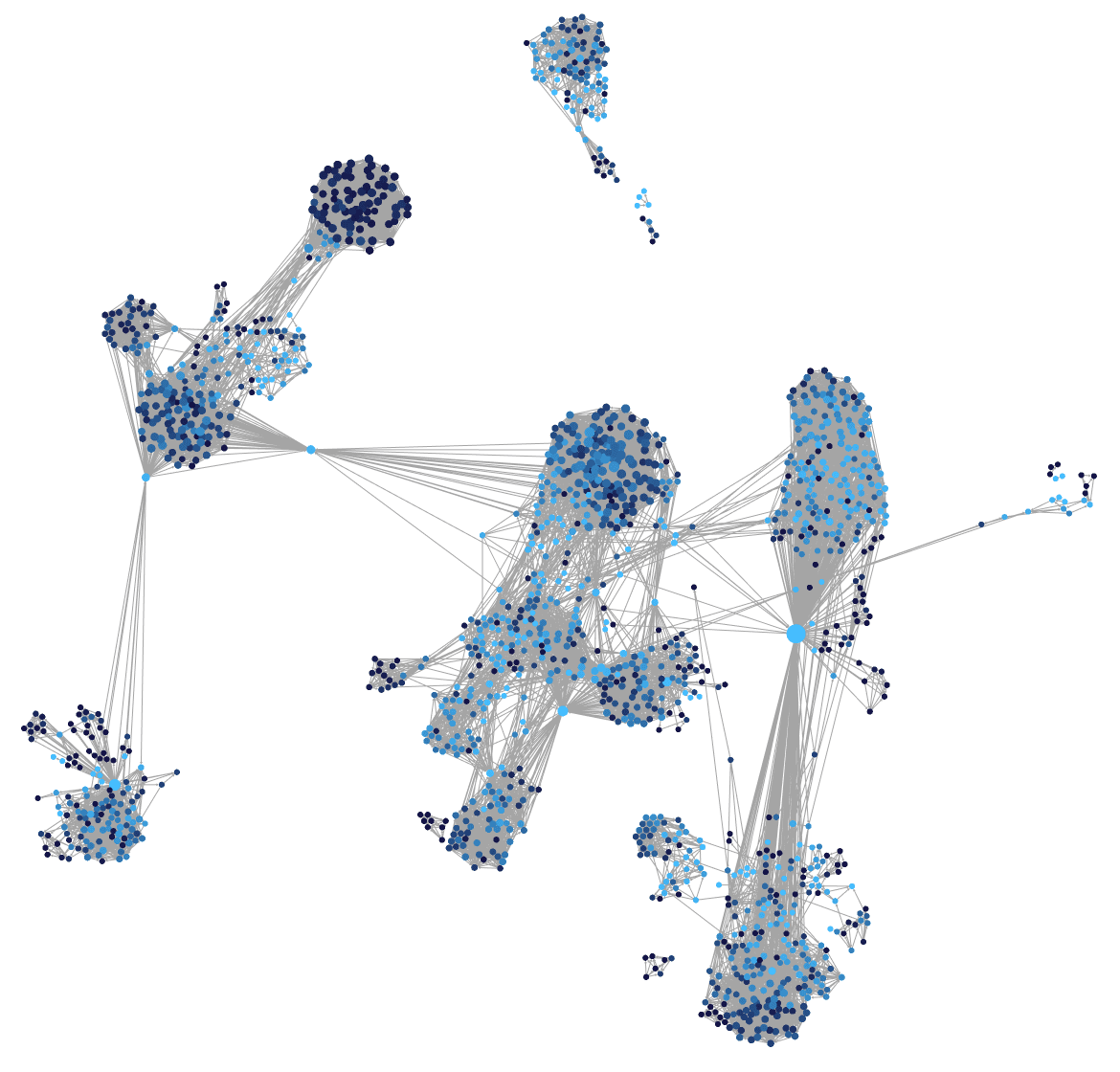}
        	\caption{\textit{RW}, $s = 0.4$}
        	\label{fig:vis-eval:facebook-rw}
    	\end{subfigure}
	\end{minipage}
	\caption{ego-Facebook graph with different sampling algorithms}
    \label{fig:vis-eval:facebook}
\end{figure*}

}
{

\begin{table*}[h]
    \scriptsize
    \centering
    \begin{tabular}{l l l r r r r r r r r r r}
    \toprule
    \multicolumn{1}{c}{\multirow{2}{*}{\textbf{Sampling}}} & \multirow{2}{*}{\textbf{Graph}} & \multirow{2}{*}{$s$} & \multicolumn{10}{c}{\textbf{Metrics}} \\
								        & & & $|V|$ & $|E|$ & $\mathcal{D}$ & $\mathcal{T}$ & $\mathcal{C_G}$ & $\mathcal{C_L}$ & $|WCC|$ & $d_{avg}$ & $d_{min}$ & $d_{max}$ \\
                                        \midrule
    \multirow{3}{*}{\textbf{Original}}  & ego-Facebook & - & 4039 & 88234 & 0.0054100 & 1612010 & 0.5191743 & 0.3027734 & 1 & 44 & 1 & 1045 \\
								        \cmidrule{2-13}
								        & ca-AstroPh & - & 18772 & 396160 & 0.0011243 & 1352117 & 0.3180837 & 0.6309897 & 290 & 43 & 1 & 504 \\
								        \cmidrule{2-13}	 
								        & web-Google & - & 875713 & 5105039 & 0.0000067 & 13391903 & 0.0552306 & 0.3544989 & 2746 & 12 & 1 & 130912 \\
                                        \midrule
    \multirow{3}{*}{\textbf{RV}}        & ego-Facebook & 0.4 & 1566 & 14202 & 0.0057986 & 102211 & 0.5438169 & 0.2779600 & 20 & 18 & 1 & 280 \\
								        \cmidrule{2-13}
								        & ca-AstroPh & 0.4 & 6728 & 63550 & 0.0014041 & 85677 & 0.3138863 & 0.5003834 & 279 & 20 & 1 & 171 \\
								        \cmidrule{2-13}	 
								        & web-Google & 0.4 & 284630 & 814286 & 0.0000101 & 855400 & 0.0562393 & 0.3135771 & 13153 & 6 & 1 & 2246 \\
                                        \midrule
    \multirow{3}{*}{\textbf{RE}}        & ego-Facebook & 0.4 & 3910 & 35347 & 0.0023123 & 103893 & 0.2084985 & 0.1123553 & 9 & 19 & 1 & 416 \\
								        \cmidrule{2-13}
								        & ca-AstroPh & 0.4 & 17976 & 158279 & 0.0004899 & 353497 & 0.2037624 & 0.2263496 & 293 & 18 & 1 & 331 \\
								        \cmidrule{2-13}	 
								        & web-Google & 0.4 & 726208 & 2040907 & 0.0000039 & 1458407 & 0.0361862 & 0.1337324 & 15763 & 6 & 1 & 2560 \\
                                        \midrule
    \multirow{3}{*}{\textbf{RVN}}       & ego-Facebook & 0.025 & 1946 & 4065 & 0.0010719 & 2701 & 0.0408925 & 0.1630864 & 5 & 5 & 1 & 202 \\
								        \cmidrule{2-13}
								        & ca-AstroPh & 0.035 & 7581 & 27818 & 0.0004842 & 4837 & 0.0297242 & 0.1871483 & 97 & 8 & 1 & 354 \\
								        \cmidrule{2-13}	 
								        & web-Google & 0.075 & 384039 & 743002 & 0.0000050 & 221799 & 0.0096453 & 0.1322633 & 11119 & 4 & 1 & 4704 \\
								        \midrule
    \multirow{3}{*}{\textbf{RW}}        & ego-Facebook & 0.4 & 1589 & 17673 & 0.0070043 & 153142 & 0.5763919 & 0.2889909 & 23 & 23 & 1 & 279 \\
								        \cmidrule{2-13}
								        & ca-AstroPh & 0.4 & 11005 & 287857 & 0.0023772 & 1060122 & 0.3320357 & 0.5829372 & 169 & 53 & 1 & 454 \\
								        \cmidrule{2-13}	 
								        & web-Google & 0.4 & 497465 & 3290006 & 0.0000133 & 9040618 & 0.0817287 & 0.3985179 & 5144 & 14 & 1 & 4050 \\
    \bottomrule
    \end{tabular}
    \caption{Metric comparison for different sampling algorithms}
    \label{tab:eval:metrics}
\end{table*}

}

\end{document}